\documentclass[twocolumn,showpacs,preprintnumbers,amsmath,amssymb,rmp]{revtex4} 

\usepackage{natbib}
\usepackage{graphicx}
\usepackage{dcolumn}
\usepackage{bm}
\usepackage{textcomp}

\begin{document}


\title{Apparatus for excitation and detection of Rydberg atoms in quantum gases}

\author{Robert L\"ow }
\email{r.loew@physik.uni-stuttgart.de}
\author{Ulrich Raitzsch}
\author{Rolf Heidemann}
\author{Vera Bendkowsky}
\author{Bj\"orn Butscher}
\author{Axel Grabowski}
\author{Tilman Pfau}%
\affiliation{ 5. Physikalisches Institut, Universit\"at Stuttgart,
Germany }

\date{\today}

\begin{abstract}
We present and characterize a versatile experimental setup which
allows for excitation and detection of Rydberg atoms in quantum
gases. The novel concept of the setup features two charged particle detectors and
eight electrical field plates inside the vacuum chamber, which
allows the detection and manipulation of Rydberg atoms. The setup
presented here is applicable to all atomic species used in the field
of quantum gases. We describe and characterize the production of
Bose-Einstein condensates, the excitation scheme into Rydberg
states, the detection of Rydberg states by field ionization followed
by ion detection and the various electric field configurations
provided by the eight field plates.

\end{abstract}

\maketitle

\section{\label{chapintroduction}Introduction}

Over the last two decades scientists have developed a huge variety
of different vacuum chamber designs serving distinctive purposes for
experiments in atom optics. An important and challenging task in
many experiments is the reliable production of Bose-Einstein
condensates \cite{Ket02,Cor02} in combination with additional
experimental tools. Depending on the physics to investigate
specialized setups have been developed explore transportable
condensates \cite{Chikkatur02}, toroidal shaped condensates
\cite{gup05}, condensates on microchips \cite{For98,Hansel01}. Other
designs featuring experimental extensions like high finesse cavities
\cite{ottl:063118} and multi-channel plates \cite{Robert01} have
been realized. Here, we present a new apparatus for the
investigation of Rydberg atoms in ultracold quantum gases. The setup
itself is based on a standard BEC-setup \cite{streed:023106}, which
is used in many atom optician laboratories over the world, though,
it has the unique capability to investigate and manipulate Rydberg
atoms in the quantum degenerate regime.

Bose-Einstein condensates are produced by trapping and cooling atoms
from a thermal atom source, usually from a  solid alkali metal.
After slowing down the velocity of the thermal atoms by several
orders of magnitude they can be trapped in a magneto-optical trap,
that is, three pairs of laser beams confining the atoms around the
center of a superposed magnetic quadrupole field. Due to scattering
of light in the magneto-optical trap it is not possible to achieve
condensation in this type of trap. Thus, the atoms have to be loaded
in a trap, in which the confining potential is not build up by near
resonant light.

One possibility to realize a conservative trapping potential is an
intense far off resonant light field (FORT). The trapping potential
is build up by the dipole force, which is due to the interaction
between the light and the induced dipole moment of the atom. Another
possibility, and by far the most common, is to trap the atoms
magnetically in a Ioffe-Pritchard type trap \cite{Ernst:98}. The
Ioffe-Pritchard trap is formed by a radial 2D quadrupole field and
an axial dipole field, which yields a radial gradient and a
superposed axial curvature. Once the atoms are tightly confined
evaporative cooling is applied to cool the atoms into the quantum
degenerated regime in which the ground state is macroscopically
occupied and the atoms form a coherent matter wave.

The combination of Bose-Einstein condensates with Rydberg atoms
opens a new field of research aiming for strong interactions and
strongly correlated mesoscopic systems. The coherence properties of
quantum degenerate gases in combination with a coherent excitation
scheme into Rydberg states yields an ideal tool box to study
decoherence processes in mesoscopic systems \cite{Heidemann07} and
for testing the preconditions for quantum computing
\cite{Jaksch:00,DiVincenzo:00}. The strong interaction between
Rydberg atoms is governed either by the van der Waals interaction,
which can induce a blockade effect in Rydberg excitation
\cite{Tong:2004,Singer:04,Heidemann07} or the anisotropic
dipole-dipole interaction. A handy feature of Rydberg atoms is the
tunability of the interaction strength by either choosing a certain
Rydberg state or by applying suitable electric fields \cite{Gal94}.

Far reaching insight in the properties of Rydberg atoms have been
achieved by experiments on thermal atomic beams \cite{Koc83}. With
the discovery of laser cooling techniques
\cite{Chu:98,Cohen:98,Phillips:98} it is now possible to extend the
experiments to so called frozen Rydberg gases
\cite{Anderson:1998,Mourachko:1998}. The name reflects the fact that
the atomic motion at temperatures in the microkelvin regime are
nearly frozen out during the typical experimental timescales of
several microseconds. The relatively high densities of several
$10^{16}$ atoms/m$^3$ privilege these systems to investigate the
interaction among Rydberg atoms. In such systems several groups have
investigated the van-der-Waals interaction
\cite{Farooqi:2003,Singer:2004,Tong:2004,Amthor:2007} and the
resonant dipole-dipole interaction
\cite{Anderson:1998,Afrousheh:2004,Vogt:2006} including the angular
dependence of the dipole-dipole interaction \cite{Carroll:2004}.
With the observation of a coherent excitation
\cite{Cubel:2005,Deiglmayr:2006} is it now possible to study the
coherence properties of mesoscopic quantum systems under the
influence of strong interactions \cite{Heidemann07}. Closely related
to the frozen Rydberg gases is the simultaneously evolving research
field on ultracold plasmas \cite{Killian:99,Robinson:2000}, which
can be studied in the same setups.

Finally, the extension of such setups to degenerate quantum gases
increases not only the parameter space of available densities and
temperatures by several orders of magnitude, but allows also for
phase sensitive measurements using a coherent matter wave with well
defined internal and external states.

Besides the ability of the apparatus to produce stable Bose-Einstein
condensates working with ultracold Rydberg atoms requires a
modification of the common setups. Especially components to apply
strong electric fields over the atomic cloud and ion or electron
detectors for the detection of Rydberg atoms are required.
Additionally a narrow-band laser system is required, to excite the
atoms into specific Rydberg states or to gain coherent control over
the excitation.

This paper is organized in the following way. We describe the vacuum
chamber in the next section, followed by a characterization of the
Bose-Einstein condensate in section \ref{chapBEC}. The sections
\ref{chapRydLaser} and \ref{chapEfield} are covering the
manipulation of the Rydberg atoms, namely, introducing the Rydberg
laser system and the electric field generation. The detection of
Rydberg atoms is presented in section \ref{chapMCP} followed by a
measurement of a Stark map in section \ref{chapStark}. In the end an
outlook is given in section \ref{chapOutlook}.

\section{\label{chapvacuum}Vacuum system}

\begin{figure}
\includegraphics[width=\linewidth]{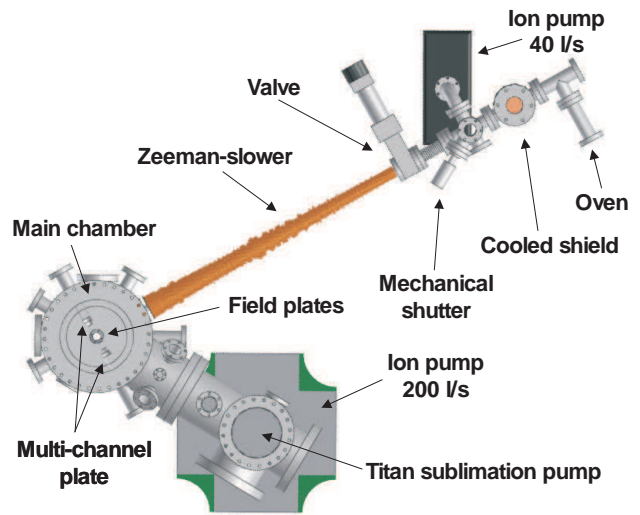}
\caption{\label{picwholechamber} Overview of the complete vacuum
system. A thermal beam of Rb-atoms is produced with an effusive oven
and slowed down with a Zeeman-slower which connects the oven part
with the main chamber. The required ultra-high vacuum in the main
chamber is accomplished with an ion pump and a titanium sublimation
pump, which can be cooled down with liquid nitrogen.  }
\end{figure}

The vacuum chamber consists mainly of two parts as shown in Fig.
\ref{picwholechamber}. The oven assembly, which is operated at high
vacuum ($10^{-7}$ mbar), delivers a thermal beam of gaseous Rubidium
atoms into the ultra-high vacuum part of the setup ($<2\cdot
10^{-11}$ mbar) for further processing.  The basic design of our
setup can be found in several experimental groups in similar
realizations and a detailed description of its basic principles can
be found in \cite{streed:023106}. At the given pressure in the main
chamber we measured a lifetime of more than 160 seconds of
magnetically trapped $^{87}Rb$ atoms.

The production of a Bose-Einstein condensate consists of several
successive experimental steps. The starting point is a
magneto-optical trap (MOT), which is in our case loaded by an atomic
beam produced by an effusive oven and slowed down by a
Zeeman-slower. Within a few seconds several $10^{10}$ atoms are
caught in the MOT. To reduce the temperature of the
magneto-optically trapped atoms to about 25-30 \textmu K we apply a
grey molasses cooling scheme for 30 ms \cite{lett:1989}. As a next
step the atoms are optically pumped to the $F=2,m_F=2$ state and
transferred into a best possible mode-matched pure magnetic trap. In
this trap the atomic cloud is cooled down within 30 s to quantum
degeneracy by forced evaporative cooling using radio-frequency
techniques. By this procedure we are able to produce Bose-Einstein
condensates with $3\cdot10^5$ atoms every 40 s.

\begin{figure*}
\includegraphics[width=\linewidth]{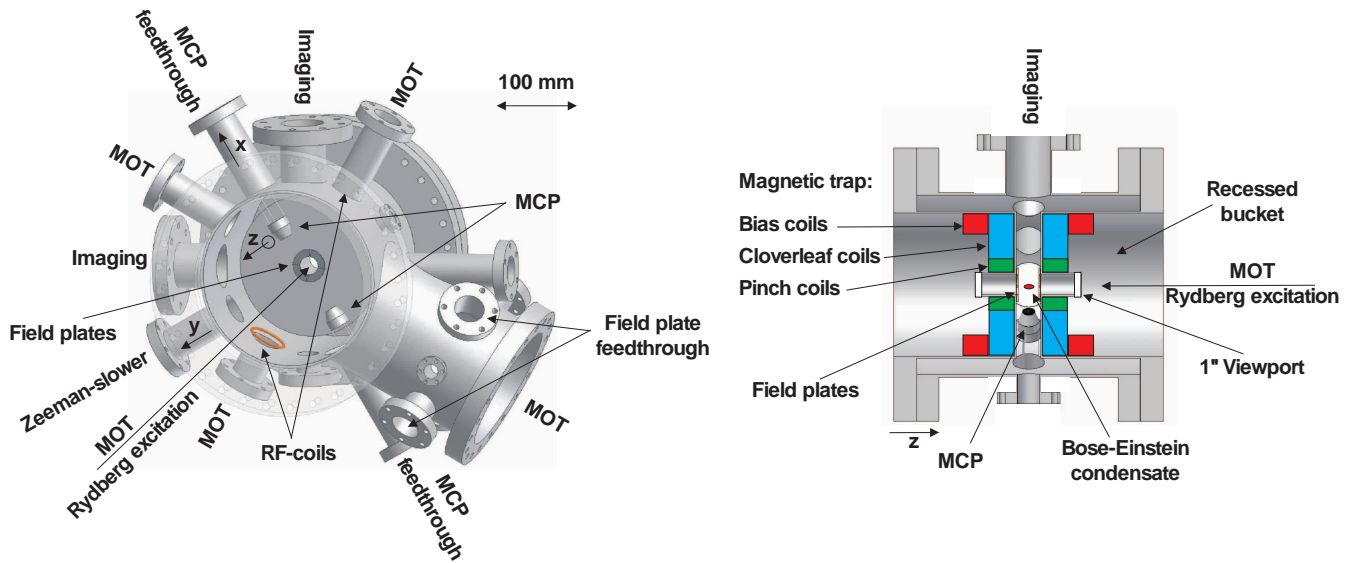}
\caption{\label{picmainchamber} Detailed view of the main chamber.
In the left part of the figure is one recessed bucket removed to
reveal the components used for electrical field manipulation and
Rydberg atom detection. The right part of the figure shows a slice
of the main chamber including both recessed buckets. Inside the
buckets, but outside the vacuum, are the coils for magnetic trapping
located. More details are given in the text.}
\end{figure*}

The main chamber depicted in Fig. \ref{picmainchamber} has to
fulfill several boundary conditions simultaneously. First of all,
good optical access in three dimensions for laser cooling is
necessary. Two further optical axes are added for imaging which are
equipped with two larger view-ports (CF 63) to obtain a better
optical resolution. The numerical apertures in both imaging axes are
0.17 which yields an optimal resolution of 5.6 \textmu m for a
wavelength of 780 nm. Altogether eleven optical view-ports (7056
glass) are available which are suitable for wavelengths ranging from
300 nm to 2.5 \textmu m.  At the same time close by magnetic coils
for magnetic trapping are necessary. We use a cloverleaf style
Ioffe-Pritchard trap \cite{Pritchard:83,streed:023106} which is
located inside two recessed bucket windows outside the vacuum. The
inner spacing of the two coil assemblies is 32 mm. The two pinch
coils of the cloverleaf trap produce an axial curvature of
$B^{\prime\prime}=0.56$ G/cm$^2$ per Ampere and the leafs a radial
gradient of $B^\prime=0.61$ G/cm per Ampere. At typical operation
conditions with 400 A in all coils and an offset field of 1.5 G we
obtain for $^{87}$Rb trapped in the $F=2,m_F=2$ state trapping
frequencies of 18 Hz axially and 250
 Hz radially.

The main goal of the setup described here is the investigation of
Rydberg atoms excited from Bose-Einstein condensates. For the
manipulation and detection of the highly excited atoms we included
eight field plates and two multi-channel plates (MCP) close to the
atoms inside the vacuum. The details of this add-ons will be
discussed in detail below. Because of the arrangement of the field
plates we had to relocate the radio-frequency coils used for
evaporative cooling further away from the atoms as can be seen in
Fig. \ref{picmainchamber} and Fig. \ref{picmcpplatte}. We use two
coils made of polyimid coated copper wires consisting each of two
loops. Remanent charge on the insulating coating can cause
disturbing electric fields if they are too close to the atoms, which
is also avoided by the larger distance of the coils. Nevertheless,
the coils produce at the position of the atoms an average magnetic
field of 5 mG (for frequencies from 1 to 30 MHz), when driven with 2
W. This is sufficient to drive the magnetic dipole transitions for
evaporative cooling.

\section{Bose-Einstein condensation}\label{chapBEC}

\begin{figure}
\includegraphics[width=\linewidth]{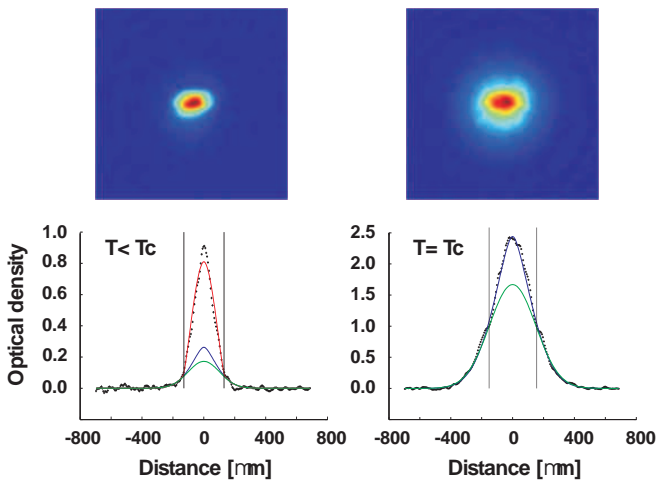}
\caption{\label{picbec} Density distribution at different
temperatures. The pictures show a central slice of two absorption
images taken after 21 ms time of flight. The figure on the right is
a thermal cloud just at the critical temperature. On the left an
atomic cloud well below $T_c$ is shown. The vertical black lines
indicate the part of the clouds which was taken for fitting the
thermal contribution. The green curve is a Gaussian fit on the wings
beyond the black lines. The blue line is the Bose enhanced
distribution, which shows that there is actually no condensate
fraction present in the right figure. Finally the red line is the
condensate fraction in Thomas-Fermi approximation.}
\end{figure}

Typically we perform evaporative cooling for 30 s on the
magnetically trapped cloud until we reach the critical temperature
$T_c$ at roughly 400 nK for the given trap parameters. The analysis
of the Bose condensed clouds is done after some time-of-flight with
the absorption imaging technique. In Fig. \ref{picbec} a purely
thermal cloud at $T\approx T_c$ is shown and a cloud at $0.7\cdot
T_c$ with a condensate fraction of 65 $\%$, both after a free
expansion of 21 ms. The temperature of the cloud is obtained by
fitting a Gaussian distribution to the wings (indicated by the
vertical lines) of the recorded density distribution. The width of
the Gaussian distribution determines then, with the knowledge of the
free expansion time, the temperature of the cloud, assuming a
point-like trapped cloud. Apparently the atomic cloud at $T=T_c$ is
not Gaussian anymore. The Bose enhancement plays already above $T_c$
a crucial role. The blue line is a fit to a Bose distribution where
the chemical potential was set to zero. From this fit we deduce the
atom number in the thermal cloud for all temperatures below $T_c$.
In a next step we subtract the fitted Bose distribution from the
data and assign the remaining part to the Bose-Einstein condensate.
The density distribution in Thomas-Fermi approximation follows a
parabolic shape for a harmonic trapping potential. With this it is
now possible to extract the condensate fraction, which is the number
of atoms in the condensed phase divided by the total atom number.
With the knowledge of the condensed atom number and the trapping
frequencies all important parameters of the atomic cloud are
determined. A trapped BEC with $3\cdot 10^5$ atoms has a radial
Thomas-Fermi radius of 3.3 \textmu m and an axially radius of 46
\textmu m. The peak density is $3.5\cdot 10^{14}$ cm$^{-3}$ and the
corresponding chemical potential is roughly 3 kHz $\cdot$ in units of Plancks constant.

\section{\label{chapRydLaser}Lasersystem for Rydberg excitation}

\begin{figure}
\includegraphics[width=\linewidth]{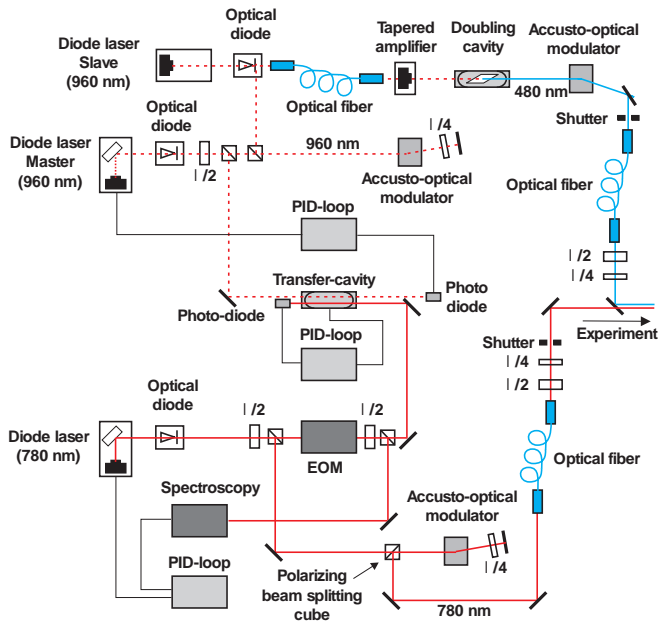}
\caption{\label{piclasersystem} Laser-system for two photon
excitation of $^{87}$Rb into Rydberg states. The red light at 780 nm
is generated by a standard external cavity diode laser system. The
production of blue light in the range of 475 nm to 483 nm is more
involved. For this purpose we use a master slave setup, where a
standard diode laser setup delivers infrared light at 960 nm which
is amplified by a tapered amplifier. A subsequent frequency doubling
cavity delivers the desired wavelength. After stabilizing a transfer
cavity to the red light at 780 nm this cavity is used to stabilize
the blue light at 480 nm with respect to the 780 nm light. The red
light at 780 nm is locked to a spectroscopy signal produced in a
Rubidium gas cell.}
\end{figure}

The excitation into Rydberg states with principal quantum numbers
ranging from n=20 up to the ionization threshold is accomplished by
a two-photon excitation scheme. For a single photon excitation one
has to use light at 297 nm, which is difficult to produce. In the
two-photon excitation scheme the first laser is tuned close to
resonance to a transition of the ground state to an intermediate
state, from which the second photon takes the atom to the desired
Rydberg state. The polarizability of the intermediate state enhances
the coupling of the ground state to the Rydberg state. By choosing a
sufficient large detuning with respect to the intermediate state one
avoids population of the intermediate state and one obtains
effectively a two level system. In our case we use as an
intermediate state the $5P_{3/2}$ state which corresponds to
wavelengths at 780 nm and 480 nm. The red light (780 nm) is produced
by a standard external cavity diode laser setup, whereas the blue
light (480 nm) is generated by frequency doubling laser light of a
tapered amplifier at 960 nm. Fig. \ref{piclasersystem} depicts the
schematic setup of the laser system. The two laser systems for the
red (780 nm) and the infrared (960 nm) light, as well all elements
for frequency stabilization are located in a separate room to
increase the stability. By this we reached an overall line-width of
the two photon excitation of about 1.5 MHz for excitation times above one millisecond.
On the timescale of our experiments of several microseconds we measure a line-width of about
130 kHz. The infrared light is brought to the laboratory by a
polarization maintaining (PM) optical fiber, passes a tapered
amplifier, a frequency doubling cavity (TA-SHG 110,
Toptica Photonics AG, Germany) and is finally guided by another
PM-fiber to the experiment. A third PM-fiber takes the light at 780
nm to the experiment.

Our setup allows us to choose
easily different laser frequencies for manifold experimental
situations. The frequency of the 780 nm light is stabilized with a
standard polarization spectroscopy method \cite{Demtroeder:02} to
any desired line or crossover of the $5S_{1/2}\rightarrow 5P_{3/2}$
transition manifold. Before it enters the vacuum chamber, it passes
an acousto-optical modulator (AOM) in a double-pass configuration.
With this AOM we can switch the light within 20 ns and tune the
frequency of the light by 200 MHz within the 3dB bandwidth. In
combination with the locking scheme we can tune the light right on
any resonance, or detune it at most by 500 MHz with respect to any
transition line of the $5S_{1/2}\rightarrow 5P_{3/2}$ manifold. The
transfer cavity is used to stabilize the 960 nm light to the already
stabilized 780 nm light \cite{Zanjani:06}. The 30 cm long cavity is
made of stainless steel and has a mode spacing of 125 MHz. Its
length can be altered by a piezo actuator, which is used to
stabilize the length of the cavity onto the transmission signal of
the 780 nm light. We evacuated the cavity to avoid changes in the
refractive index of the air due to atmospheric pressure and
humidity. The master laser at 960 nm can now be stabilized to any
mode of the cavity in steps of 125 MHz. A subsequent AOM in double
pass configuration allows us to scan the infrared light by 300 MHz,
which corresponds to 600 MHz of the frequency doubled blue light at
480 nm. To scan a larger frequency region than available by the AOM,
we control the grating of the infrared diode laser directly and scan
by this the frequency without mode hops for more than 12 GHz in the
blue light. To calibrate the frequency of the scanned region we
simultaneously record the signal from the stabilized Fabry Perot
resonator.

The two laser beams for the two-photon excitation are spatially
overlapped before entering the vacuum chamber at a dichroic mirror.
The overlapped beams were aligned parallel to the z-axis, which is
also the quantization axis of the magnetic trap. The polarization of
both beams was set to $\sigma^+$ respectively $\sigma^-$ to maintain
the magnetic moment of the atom and avoid frequency shifts due to
magnetic fields. At the position of the atoms the 1/e$^2$ radius of
the 780 nm light was set to 550 \textmu m and the one at 480 nm to
35 \textmu m. The maximum available laser power in the blue light (480 nm) is
about 55 mW  at the position of the atoms. In most experiments the power of the 780 nm light
is reduced well below one milliwatt to avoid excitation into the $5P_{3/2}$ state.
At an detuning of 480 MHz (see Fig. \ref{picexcite}) and a typical laser power
of e.g. 50 \textmu W the spontaneous scattering rate reduces to below one 1 kHz. The effective
two photon Rabi frequency at this setting is 250 kHz.
Another important aspect the uniformity of the illumination of the atoms. An atomic cloud
at e.g. 3.4 \textmu K confined in our magnetic
trap at an offset field of 0.89 G, as used in \cite{Heidemann07}, has a Gaussian shape with a radial
width of $\sigma_\rho=8.6$ \textmu m. At this parameters
85 \% of the atoms experience at least 80 \% of the maximum
two photon Rabi frequency.



\section{\label{chapEfield}Electric Field Plates}

\begin{figure}
\includegraphics[width=\linewidth]{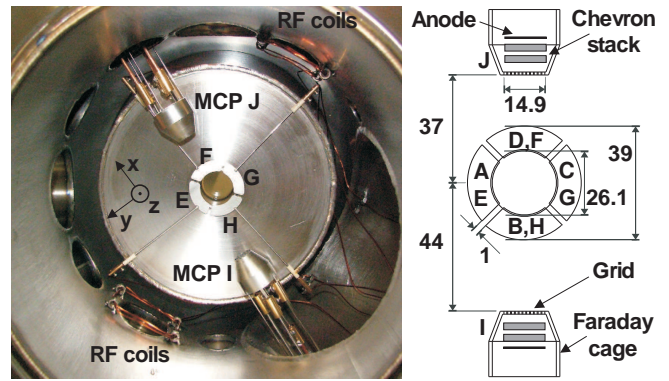}
\caption{\label{picmcpplatte} Electric field plates (A-H) and
Faraday cages ($I$ and $J$) for the multi-channel plates. Four field
plates are glued onto each of the resessed buckets,  such that plate
$A$, $B$, $C$, and $D$ lie vis-a-vis to the plates $E$, $F$, $G$ and
$H$. The inner distance between the plates is 25 mm. All dimensions
given in the figure are in millimeters. The MCPs were located as
close as possible to the center of the vacuum chamber without
loosing any optical access, which resulted in the two different
distances. }
\end{figure}

\begin{figure}
\includegraphics[width=\linewidth]{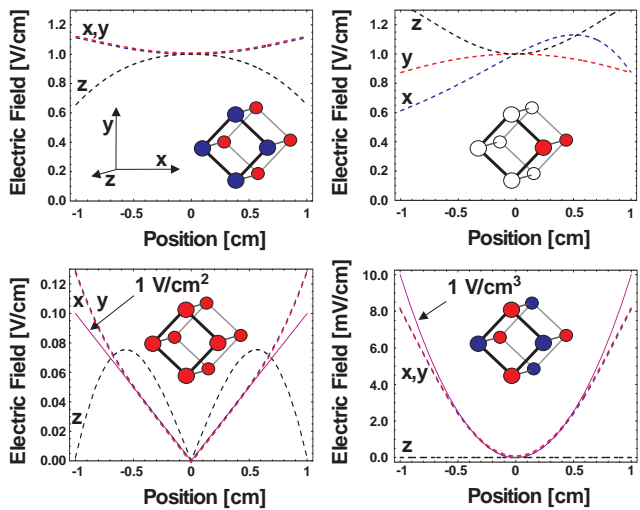}
\caption{\label{picefelder} Simulation of four different electric
field configurations produced by an asymmetric octopole. The field
plates are represented by point charges located at the corners of a
cuboid with an aspect ratio of x:y:z=15:15:14 (mm). This simple
geometry gives an excellent agreement with the electric fields of
the real geometry, which have also been simulated by an elaborate
finite element method. All four graphs show the absolute value of
the calculated electric field along the coordinate axes as defined
in Fig. \ref{picmainchamber} and Fig. \ref{picmcpplatte}. The upper
left graph shows the realization of a preferably constant field by
setting the charges A, B, C and D to a charge -Q (blue) and E, F, G and H
accordingly to +Q (red). The upper right graph shows a simplified
implementation of a constant electric field by setting charges B and
H to +Q, which was used in our measurements in section
\ref{chapStark}. On the lower left side is the realization of a
gradient field along all three axes is shown (by setting all charges
to the same value), which could be used in order to excite atoms
space selective into Rydberg states. Finally the graph on the lower
right shows an electric field distribution with a curvature in two
directions, which is realized by charging the plates in an alternate
pattern.}
\end{figure}

The high sensitivity of Rydberg atoms to electric fields opens the
possibility to manipulate the internal states of the Rydberg atoms
by field plates \cite{Adam:03}. To produce electric field
configurations as versatile as possible, we installed eight field
plates close to the atoms. The spatial arrangement can be seen in
Fig. \ref{picmcpplatte} and Fig. \ref{picmainchamber}. Each of these
plates can be addressed individually, which allows us to generate
nearly any field configuration starting from constant fields in
arbitrary directions, to gradient fields, quadrupolar fields,
hyperbolic concave saddles or hyperbolic convex fields with a
positive curvature in all three dimensions. Some realizations are
shown in Fig. \ref{picefelder}.

The eight field plates are made of stainless steel with a thickness
of 0.5 mm. They are glued with 1 mm thick ceramic spacers onto the
resessed buckets. The dimensions of the spacers have to be small
enough, that they are completely hidden behind the field plates from
the viewpoint of the atoms. Any insulating surface can accumulate
charge and falsify the desired field configuration. To charge the
field plates they are spot welded to a stainless steel wire, which
are radially led outwards as can be seen in Fig.
\ref{picmainchamber}. At the edge of the resessed bucket the wires
are fixed in position by short ceramic tubings, which are also glued
to the buckets and are subsequently connected to capton insulated
copper wires. These copper wires are then finally connected to one
of the fourfold high voltage feedthroughs. To avert breakthroughs
inside the chamber induced by sharp edges we rounded off all four
edges of each plate with a radius of 1.5 mm. Finally, we etched and
electro-polished all field plates including the spot welded wires to
burnish also small spikes. The polishing was done in a acid bath
consisting of one part of 96\% sulfuric acid, two parts of 85\%
phosphoric acid and six parts of distilled water. After two minutes
at a current of 5 A about 70 \textmu m of stainless steel from the
plates was removed and they exhibited a semi gloss surface. After
installation of the field plates and evacuating the chamber we
measured no current leakage up to 3000 Volts for all plates.

During the experimental process it is necessary to switch the
applied voltages within short times. To do so we use bipolar high
voltage switches (HTS-6103 GSM, Behlke Electronic GmbH, Germany),
which have an intrinsic rise-time of 60 ns. The push-pull circuit of
the switch has to be adjusted to match the capacitive load of 50 pF
of each field plate as well the 300 pF load of the high voltage coax
cable, which connects the switch to the high voltage feedthroughs.

The electric field distributions generated by the field plates and
the Faraday cages of the MCPs have been calculated for several
applications by finite element methods. This method is quite time
consuming but gives good results for a given set of voltages applied
to the 8 field plates and the two Faraday cages. But it is useless,
when starting from a desired field distribution and asking for the
required voltages. To do so, we use a simplified model consisting of
point charges sitting at the corners of a cuboid. By expansion of
the electric field into a Taylor series one can generate a set of
linear equations, which can be used to calculate the voltages for a
given electric field as shown in Fig. \ref{picefelder}.

\section{\label{chapMCP}Detection of Rydberg states}

For a high detection sensitivity of Rydberg atoms, we installed two
MCPs (Type B012VA, El-Mul Technologies Ltd, Israel) inside the
vacuum chamber. After field ionization of the Rydberg atoms, we use
one MCP to detect the ions. The second MCP is designated to detect
simultaneously the electrons. To improve the amplification even
further we use MCPs in a Chevron configuration, which consist of two
successive glass plates with a small spacing in-between. The
electron current arriving at the anode is converted by a large
resistor to a voltage and then amplified by a homebuild circuit
including a low noise operation amplifier. The whole MCP Chevron
assembly is boxed into a Faraday cage in order to shield the atoms
in the center of the chamber from the biased front side, typically
charged with -2000 V. The Faraday cage is closed at the front by a
grid with a diameter of 12 mm and a transmittance of 85\%. The
active area of the MCP front side has a diameter of 8.5 mm.

To detect the Rydberg atoms with an MCP one has at first to field
ionize the excited atoms. This can be done by a large enough
electric field, which is in the case of an 43S$_{1/2}$ state about
160 V/cm. In our case we want to detect the ions and one has to
provide besides a sufficient field strength also a suitable electric
field distribution which guides the ions into the MCP $J$.

Usually the magnetic fields of the trapping potential are still
switched on when the ions move towards the MCP. The combination of
electric and magnetic fields provoke a drift on the ions according
to the force $\vec{F}=q(\vec{E}+\vec{v}\times\vec{B})$. An
estimation for the given experimental situation shows that the drift
is in our case of the order of 1 mm and by this well below the
aperture of the MCP of 8.5 mm.

\begin{figure}
\includegraphics[width=\linewidth]{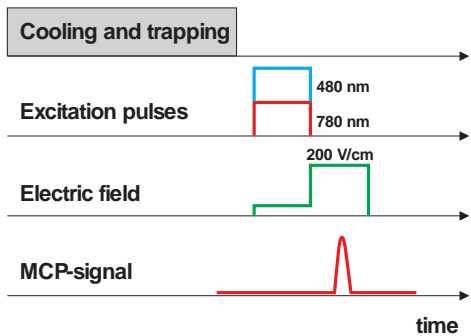}
\caption{\label{picsequenz} Experimental sequence of Rydberg
excitation. After the cooling and trapping steps are both lasers for
the two photon excitation switched on simultaneously. The pulses can
be as short as 200 ns. During laser excitation we apply an electric
field over the atoms to induce a Stark shift. It is advantageous to
apply a small electrical field also in other experiments to extract
unwanted ions from the atomic cloud. To detect the generated Rydberg
atoms we increase the electric field above 200 V/cm to field ionize
them. After a short time of flight (few \textmu s) the ions arrive
at the MCP and generate after amplification a pulse on the anode
which is recorded by a computer for further processing.}
\end{figure}

We calibrated the MCP ion signal by monitoring the losses in a cold
atomic cloud due to Rydberg excitation and the corresponding voltage
signal on the anode. After amplification, which is the same for the
subsequent experiments, we acquire a signal of 1 Vs per $3.65\cdot
10^{10}$ atoms. In principle one could distinguish between single
ion events, but the noise level of the signal limits our minimum
sensitivity to about one hundred ions.

\section{\label{chapRydMT}Rydberg excitation of magnetically trapped atoms}

\begin{figure}
\includegraphics[width=\linewidth]{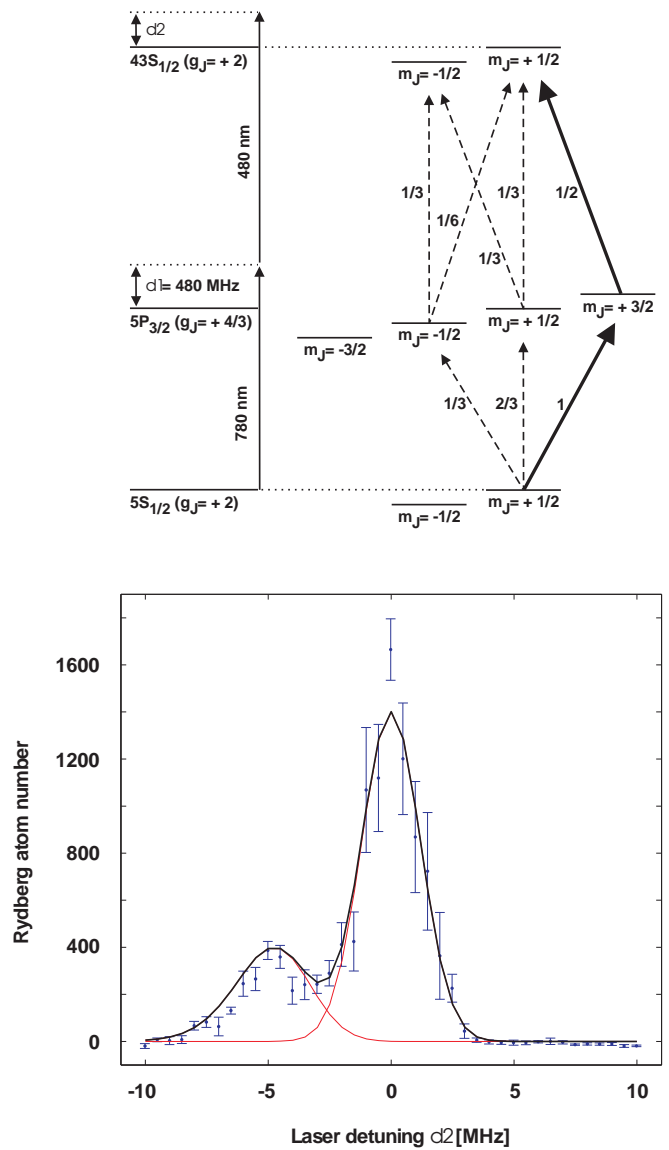}
\caption{\label{picexcite} Excitation of Rydberg atoms into the
$43S_{1/2}$ state from a cloud of magnetically trapped atoms with a
temperature of 15 $\mu$K. The upper part of the figure shows the two
photon excitation scheme and the different pathways into the Rydberg
states in the presence of a magnetic field as described in the text.
The numbers next to the arrows indicate the individual transitions
strengths. The measurement of the Rydberg excitation spectrum shown
below features as expected two distinct peaks. The peak on the right
corresponds to the in first order magnetically independent
transition $5S_{1/2},m_J=1/2\rightarrow\ 43S_{1/2},m_J=1/2$. The
underlying red line is a Gaussian fit to the data with a width of
1.5 MHz. The peak on the left raises by the excitation into the
$43S_{1/2},m_J=-1/2$ state, which is sensitive to the magnetic
field. The position and the shape of this peak is determined by the
magnetic field distribution across the atomic cloud, the
corresponding polarization pattern seen by the atoms, the
temperature of the cloud, the excitation line-width and the different possible excitation
paths. This has been calculated for the given parameters and the
only remaining fitting parameter of the red curve is the height.
Finally, the black curve is the sum of the two individual red
fitting curves.}
\end{figure}

As a first performance test we excited Rydberg atoms in a
magnetically trapped cloud of evaporatively cooled atoms. A
schematic view of the experimental sequence is shown in Fig.
\ref{picsequenz}. The ultracold cloud consisted of $4\cdot 10^7$
atoms at a temperature of 15 $\mu$K confined in a cigar shaped harmonic
trapping potential with an axial trapping frequency (along the
z-axis) of 18 Hz and an radial trapping frequency of 310 Hz. The
magnetic offset field at the center of the trapping potential was
set to 1.0 Gauss and the cloud is spin polarized in the $F=2,m_F=2$
ground state with respect to the quantization axis given by the
magnetic field. For highly excited states the hyperfine coupling is
significantly decreased and we remain with the coupling of the electron spin to the
orbital momentum to a total momentum $J$ as shown in figure
\ref{picexcite}. We want to excite the atoms into the $43S_{1/2}$
state which exhibits the same magnetic moment as the ground state
and the transition is insensitive to magnetic fields. The laser light of
both beams is circular polarized ($\sigma^+$ for 780 nm and
$\sigma^-$ for 480 nm) with respect to the z-axis (bold arrows in
Fig. \ref{picexcite}). Nevertheless alters the direction of the
magnetic field lines of a magnetic trapping potential when moving
away from the center, which leads to a space dependent angle between
the k-vector of the excitation lasers and the quantization axis. By
this it is also possible to excite atoms into other excited states
by admixtures of other polarizations.

Fig. \ref{picexcite} shows the measurement of an excitation
spectrum. To avoid a too large excited state fraction and with this
unwanted Rydberg-Rydberg interactions effects, we reduced the
excitation time to 1 \textmu s. During the excitation pulse we applied a
small electric field of 2 V/cm to extract accidental ions. After
excitation we immediately field ionized the Rydberg atoms and
detected the ions with the multichannel-plate as described in
section \ref{chapMCP}.

In an additional experiment, we measured the radiative lifetime of the $43S_{1/2}$ state
at a temperature of 20 \textmu K and a atomic peak density of $5\cdot 10^{12}$ cm$^{-3}$.
We measured a lifetime of 99$\pm$15 \textmu s, which is nicely in accordance with the expected lifetime
of 99 \textmu s. The theoretical value has been calculated with the help of the
quantum defects of $^{87}$Rb \cite{Gal94}.
This shows that the lifetime is not reduced by collisions at such high densities. One would expect a
reduction of the lifetime by a factor of two due to blackbody radiation at 300 K \cite{Gal94}, which is not seen in
our experiment. This is most likely due to the strongly reduced spectral density of modes
inside the metallic vacuum chamber \cite{Kolokolov:74}.

\section{\label{chapStark}Electric field calibration by measuring the Stark shift of the 43S$_{\text{1/2}}$ state}

\begin{figure}
\includegraphics[width=\linewidth]{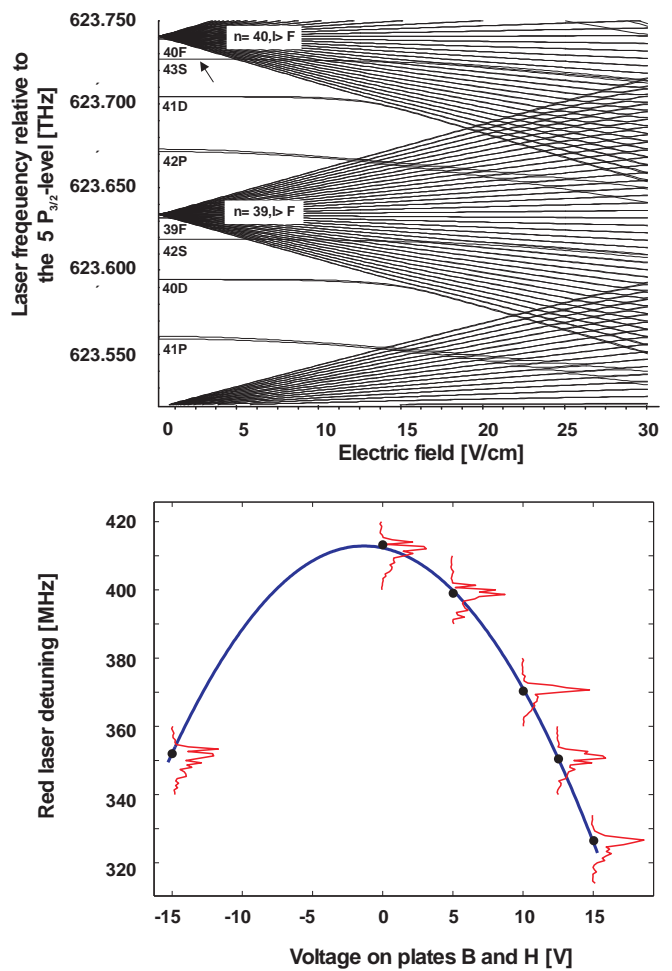}
\caption{\label{picstarkmap} The upper graph shows a theoretical
calculation of the Stark shift in the vicinity of $n=40$. The arrow
indicates the Stark shift of the 43S state which we used in our
measurements shown in the graph below. The solid blue line is a
parabolic fit since a quadratic Stark shift is expected. The red
curves show the corresponding ion signal detected on the MCPs. The
center of masses are represented by the black dots. }
\end{figure}

To examine the electric fields generated by the field plates, we
make use of the quadratic Stark effect exhibited by the 43S$_{1/2}$
state. The energy shift $\Delta W$ (in units of Plancks constant) of this state located in an
electric field $E$ is given by

\begin{equation}
\Delta W=-\frac{1}{2} \alpha E^2
\end{equation}

with $\alpha/2=8.06$ MHz/(V/cm)$^2$. This value has been calculated
by a first principles calculation of the Stark map as shown in Fig.
\ref{picstarkmap}.

The electric field for the Stark shift was produced by voltages on
the field plates B and H, which were tuned from -15 V to +15 V. All
other field plates as well the cage of the MCP I were set to ground.
The simulation of the emerging field gives in the geometric center a
value of 0.14 V/cm per applied unit of voltage on field plate B and
H. The orientation of the field is parallel to the x-axis. The
intrinsic asymmetric configuration of the field exhibits
additionally a gradient along the x-axis of 0.1 V/cm$^2$ per applied
unit volt. This field is superposed with an additional extraction
field of the Faraday cage of the MCP J, which was charged to -15 V. The
calculated field remaining from the cage at the center is 0.2 V/cm
plus a gradient of 2 V/cm$^2$. For detection of the Rydberg atoms we
switch after excitation the voltage on plate B and H to +1000 V,
which is sufficient for field ionization. The field configuration
with the plates B and H at a positive voltage and the Faraday cage
at -15~V drags the positive ions towards the MCP for detection.

As a sample for excitation we used an evaporatively cooled atomic
cloud confined in the magnetic trap. At a temperature of 40 \textmu
K the cloud is in the harmonic regime of the magnetic trap with a
radial width of about 40 \textmu m. For such small clouds one can
neglect the broadening of the spectroscopic lines due to
inhomogeneties of the electric field. For
the given parameters above, the broadening is less than two percent
of the magnitude of the Stark shift energy.

Fig. \ref{picstarkmap} shows the result on the measured Stark shift.
The parabolic fit (blue line) is shifted by 1.34 V with respect to 0
V, which is caused by the additional offset field of the Faraday
cage. At 0 V one gets a Stark shift of 0.58 MHz which corresponds to
an offset field of 0.27 V/cm (calculated 0.2 V/cm). The curvature of
the fitted parabola can be used to calibrate the electric field for
the applied voltages at field plate B and H. To achieve 1 V/cm at
the position of the atoms one has to apply 5 V at both field plates.
The expected value from the theoretical calculation is 7.2 V. Both
deviations from the theoretical can be explained if the atomic cloud
is shifted along the x-axis by a few millimeters away from the
geometric center. More likely are imprecisions of the numerical
calculations, since the parametrization of the full vacuum chamber
geometry is not feasible.

The spectroscopic lines (plotted in red) exhibit a broadening which
is not independent of the applied electric field. There are two
possible paths for the two photon transition $5S_{1/2}\rightarrow
5P_{3/2}\rightarrow 43S_{1/2}$. The desired transition ends in the
$j=+1/2$ state of the $43S_{1/2}$ manifold, which is the only one
allowed by the chosen polarizations. Nevertheless, is the direction
of the magnetic field adjacent to the trapping center not parallel
to the z-axis and atoms can also be transferred to the $j=-1/2$
which is not anymore insensitive to the magnetic field. The $g_j$
factor of the Rydberg state is 2 and the total change in magnetic
momentum is 2 $\mu_B$. A calculation for the given parameters
results in a broadening of about 3 MHz, which explains the observed
linewidths very well.

\section{\label{chapOutlook} Conclusion}

In this article, we have presented details of designing a setup
which combines quantum degenerate gases with Rydberg atoms. The
capability of coherent excitation into specific Rydberg states and
their efficient detection with multichannel-plates establishes a new
framework to study coherence properties of mesoscopic quantum
systems. Especially the tunability of the interaction strength and
type among the Rydberg atoms by either choosing a specific Rydberg state
or by applying electric fields
make this a valuable system for many
tasks. The first realization of
coherent and collective excitation in a
mesoscopic quantum system in the strong coupling regime has been
achieved with this setup \cite{Heidemann07}. There exist manifold
options for further investigations. This could be studies on the coherence
properties of nonclassical states, the dependence on system size and
its geometry, signatures beyond mean-field theories and the
feasibility of such systems for quantum information.


\section{Acknowledgments}

Funding of this setup was provided by the Deutsche Physikalischen
Gesellschaft in the Schwerpunktprogramm (SPPP 1116) and by the
Sonderforschungsbereich SFB/TR 21. The Rydberg laser system was
partly funded by the Landesstiftung Baden-W\"urttemberg. The authors
thank J\"org Bauer, Christian Kuke, Johannes Nold,
Eva Kuhnle and Helmar Bender for their contributions in setting up this experiment.


\begin{thebibliography}{38}
\expandafter\ifx\csname natexlab\endcsname\relax\def\natexlab#1{#1}\fi
\expandafter\ifx\csname bibnamefont\endcsname\relax
  \def\bibnamefont#1{#1}\fi
\expandafter\ifx\csname bibfnamefont\endcsname\relax
  \def\bibfnamefont#1{#1}\fi
\expandafter\ifx\csname citenamefont\endcsname\relax
  \def\citenamefont#1{#1}\fi
\expandafter\ifx\csname url\endcsname\relax
  \def\url#1{\texttt{#1}}\fi
\expandafter\ifx\csname urlprefix\endcsname\relax\def\urlprefix{URL }\fi
\providecommand{\bibinfo}[2]{#2}
\providecommand{\eprint}[2][]{\url{#2}}

\bibitem[{\citenamefont{{Afrousheh}}
  \emph{et~al.}(2004)\citenamefont{{Afrousheh}, {Bohlouli-Zanjani}, {Vagale},
  {Mugford}, {Fedorov}, and {Martin}}}]{Afrousheh:2004}
\bibinfo{author}{\bibnamefont{{Afrousheh}}, \bibfnamefont{K.}},
  \bibinfo{author}{\bibfnamefont{P.}~\bibnamefont{{Bohlouli-Zanjani}}},
  \bibinfo{author}{\bibfnamefont{D.}~\bibnamefont{{Vagale}}},
  \bibinfo{author}{\bibfnamefont{A.}~\bibnamefont{{Mugford}}},
  \bibinfo{author}{\bibfnamefont{M.}~\bibnamefont{{Fedorov}}}, and
  \bibinfo{author}{\bibfnamefont{J.~D.} \bibnamefont{{Martin}}},
  \bibinfo{year}{2004}, \bibinfo{journal}{Physical Review Letters}
  \textbf{\bibinfo{volume}{93}}(\bibinfo{number}{23}), \bibinfo{pages}{233001}.

\bibitem[{\citenamefont{{Amthor}} \emph{et~al.}(2007)\citenamefont{{Amthor},
  {Reetz-Lamour}, {Westermann}, {Denskat}, and
  {Weidem{\"u}ller}}}]{Amthor:2007}
\bibinfo{author}{\bibnamefont{{Amthor}}, \bibfnamefont{T.}},
  \bibinfo{author}{\bibfnamefont{M.}~\bibnamefont{{Reetz-Lamour}}},
  \bibinfo{author}{\bibfnamefont{S.}~\bibnamefont{{Westermann}}},
  \bibinfo{author}{\bibfnamefont{J.}~\bibnamefont{{Denskat}}}, and
  \bibinfo{author}{\bibfnamefont{M.}~\bibnamefont{{Weidem{\"u}ller}}},
  \bibinfo{year}{2007}, \bibinfo{journal}{Physical Review Letters}
  \textbf{\bibinfo{volume}{98}}(\bibinfo{number}{2}), \bibinfo{pages}{023004}.

\bibitem[{\citenamefont{{Anderson}}
  \emph{et~al.}(1998)\citenamefont{{Anderson}, {Veale}, and
  {Gallagher}}}]{Anderson:1998}
\bibinfo{author}{\bibnamefont{{Anderson}}, \bibfnamefont{W.~R.}},
  \bibinfo{author}{\bibfnamefont{J.~R.} \bibnamefont{{Veale}}}, and
  \bibinfo{author}{\bibfnamefont{T.~F.} \bibnamefont{{Gallagher}}},
  \bibinfo{year}{1998}, \bibinfo{journal}{Physical Review Letters}
  \textbf{\bibinfo{volume}{80}}, \bibinfo{pages}{249}.

\bibitem[{\citenamefont{{Bohlouli-Zanjani}}
  \emph{et~al.}(2006)\citenamefont{{Bohlouli-Zanjani}, {Afrousheh}, and
  {Martin}}}]{Zanjani:06}
\bibinfo{author}{\bibnamefont{{Bohlouli-Zanjani}}, \bibfnamefont{P.}},
  \bibinfo{author}{\bibfnamefont{K.}~\bibnamefont{{Afrousheh}}}, and
  \bibinfo{author}{\bibfnamefont{J.~D.~D.} \bibnamefont{{Martin}}},
  \bibinfo{year}{2006}, \bibinfo{journal}{Review of Scientific Instruments}
  \textbf{\bibinfo{volume}{77}}, \bibinfo{pages}{3105}.

\bibitem[{\citenamefont{{Carroll}} \emph{et~al.}(2004)\citenamefont{{Carroll},
  {Claringbould}, {Goodsell}, {Lim}, and {Noel}}}]{Carroll:2004}
\bibinfo{author}{\bibnamefont{{Carroll}}, \bibfnamefont{T.~J.}},
  \bibinfo{author}{\bibfnamefont{K.}~\bibnamefont{{Claringbould}}},
  \bibinfo{author}{\bibfnamefont{A.}~\bibnamefont{{Goodsell}}},
  \bibinfo{author}{\bibfnamefont{M.~J.} \bibnamefont{{Lim}}}, and
  \bibinfo{author}{\bibfnamefont{M.~W.} \bibnamefont{{Noel}}},
  \bibinfo{year}{2004}, \bibinfo{journal}{Physical Review Letters}
  \textbf{\bibinfo{volume}{93}}(\bibinfo{number}{15}), \bibinfo{pages}{153001}.

\bibitem[{\citenamefont{Chikkatur} \emph{et~al.}(2002)\citenamefont{Chikkatur,
  Shin, Leanhardt, Kielpinski, Tsikata, Gustavson, Pritchard, and
  Ketterle}}]{Chikkatur02}
\bibinfo{author}{\bibnamefont{Chikkatur}, \bibfnamefont{A.~P.}},
  \bibinfo{author}{\bibfnamefont{Y.}~\bibnamefont{Shin}},
  \bibinfo{author}{\bibfnamefont{A.~E.} \bibnamefont{Leanhardt}},
  \bibinfo{author}{\bibfnamefont{D.}~\bibnamefont{Kielpinski}},
  \bibinfo{author}{\bibfnamefont{E.}~\bibnamefont{Tsikata}},
  \bibinfo{author}{\bibfnamefont{T.~L.} \bibnamefont{Gustavson}},
  \bibinfo{author}{\bibfnamefont{D.~E.} \bibnamefont{Pritchard}}, and
  \bibinfo{author}{\bibfnamefont{W.}~\bibnamefont{Ketterle}},
  \bibinfo{year}{2002}, \bibinfo{journal}{Science} (\bibinfo{number}{296}),
  \bibinfo{pages}{2193}.

\bibitem[{\citenamefont{Chu}(1998)}]{Chu:98}
\bibinfo{author}{\bibnamefont{Chu}, \bibfnamefont{S.}}, \bibinfo{year}{1998},
  \bibinfo{journal}{Rev. Mod. Phys.} \textbf{\bibinfo{volume}{70}},
  \bibinfo{pages}{685}.

\bibitem[{\citenamefont{Cohen-Tannoudji}(1998)}]{Cohen:98}
\bibinfo{author}{\bibnamefont{Cohen-Tannoudji}, \bibfnamefont{C.~N.}},
  \bibinfo{year}{1998}, \bibinfo{journal}{Rev. Mod. Phys.}
  \textbf{\bibinfo{volume}{70}}, \bibinfo{pages}{707}.

\bibitem[{\citenamefont{Cornell and Wieman}(2002)}]{Cor02}
\bibinfo{author}{\bibnamefont{Cornell}, \bibfnamefont{E.~A.}}, and
  \bibinfo{author}{\bibfnamefont{C.~E.} \bibnamefont{Wieman}},
  \bibinfo{year}{2002}, \bibinfo{journal}{Rev. Mod. Phys.}
  (\bibinfo{number}{74}), \bibinfo{pages}{875}.

\bibitem[{\citenamefont{{Cubel}} \emph{et~al.}(2005)\citenamefont{{Cubel},
  {Teo}, {Malinovsky}, {Guest}, {Reinhard}, {Knuffman}, {Berman}, and
  {Raithel}}}]{Cubel:2005}
\bibinfo{author}{\bibnamefont{{Cubel}}, \bibfnamefont{T.}},
  \bibinfo{author}{\bibfnamefont{B.~K.} \bibnamefont{{Teo}}},
  \bibinfo{author}{\bibfnamefont{V.~S.} \bibnamefont{{Malinovsky}}},
  \bibinfo{author}{\bibfnamefont{J.~R.} \bibnamefont{{Guest}}},
  \bibinfo{author}{\bibfnamefont{A.}~\bibnamefont{{Reinhard}}},
  \bibinfo{author}{\bibfnamefont{B.}~\bibnamefont{{Knuffman}}},
  \bibinfo{author}{\bibfnamefont{P.~R.} \bibnamefont{{Berman}}}, and
  \bibinfo{author}{\bibfnamefont{G.}~\bibnamefont{{Raithel}}},
  \bibinfo{year}{2005}, \bibinfo{journal}{\pra}
  \textbf{\bibinfo{volume}{72}}(\bibinfo{number}{2}), \bibinfo{pages}{023405}.

\bibitem[{\citenamefont{{Deiglmayr}}
  \emph{et~al.}(2006)\citenamefont{{Deiglmayr}, {Reetz-Lamour}, {Amthor},
  {Westermann}, {de Oliveira}, and {Weidem{\"u}ller}}}]{Deiglmayr:2006}
\bibinfo{author}{\bibnamefont{{Deiglmayr}}, \bibfnamefont{J.}},
  \bibinfo{author}{\bibfnamefont{M.}~\bibnamefont{{Reetz-Lamour}}},
  \bibinfo{author}{\bibfnamefont{T.}~\bibnamefont{{Amthor}}},
  \bibinfo{author}{\bibfnamefont{S.}~\bibnamefont{{Westermann}}},
  \bibinfo{author}{\bibfnamefont{A.~L.} \bibnamefont{{de Oliveira}}}, and
  \bibinfo{author}{\bibfnamefont{M.}~\bibnamefont{{Weidem{\"u}ller}}},
  \bibinfo{year}{2006}, \bibinfo{journal}{Optics Communications}
  \textbf{\bibinfo{volume}{264}}, \bibinfo{pages}{293}.

\bibitem[{\citenamefont{Demtr\"oder}(2002)}]{Demtroeder:02}
\bibinfo{author}{\bibnamefont{Demtr\"oder}, \bibfnamefont{W.}},
  \bibinfo{year}{2002}, \emph{\bibinfo{title}{Laser Spectroscopy}}
  (\bibinfo{publisher}{Springer Verlag}), \bibinfo{note}{2nd edition}.

\bibitem[{\citenamefont{DiVincenzo}(2000)}]{DiVincenzo:00}
\bibinfo{author}{\bibnamefont{DiVincenzo}, \bibfnamefont{D.~P.}},
  \bibinfo{year}{2000}, \bibinfo{journal}{Fortschr. Phys.}
  \textbf{\bibinfo{volume}{48}}, \bibinfo{pages}{771}.

\bibitem[{\citenamefont{Ernst} \emph{et~al.}(1998)\citenamefont{Ernst, Marte,
  Schreck, Schuster, and Rempe}}]{Ernst:98}
\bibinfo{author}{\bibnamefont{Ernst}, \bibfnamefont{U.}},
  \bibinfo{author}{\bibfnamefont{A.}~\bibnamefont{Marte}},
  \bibinfo{author}{\bibfnamefont{F.}~\bibnamefont{Schreck}},
  \bibinfo{author}{\bibfnamefont{J.}~\bibnamefont{Schuster}}, and
  \bibinfo{author}{\bibfnamefont{G.}~\bibnamefont{Rempe}},
  \bibinfo{year}{1998}, \bibinfo{journal}{Euro. Phys. Lett.}
  \textbf{\bibinfo{volume}{41}}, \bibinfo{pages}{1}.

\bibitem[{\citenamefont{{Farooqi}} \emph{et~al.}(2003)\citenamefont{{Farooqi},
  {Tong}, {Krishnan}, {Stanojevic}, {Zhang}, {Ensher}, {Estrin}, {Boisseau},
  {C{\^o}t{\'e}}, {Eyler}, and {Gould}}}]{Farooqi:2003}
\bibinfo{author}{\bibnamefont{{Farooqi}}, \bibfnamefont{S.~M.}},
  \bibinfo{author}{\bibfnamefont{D.}~\bibnamefont{{Tong}}},
  \bibinfo{author}{\bibfnamefont{S.}~\bibnamefont{{Krishnan}}},
  \bibinfo{author}{\bibfnamefont{J.}~\bibnamefont{{Stanojevic}}},
  \bibinfo{author}{\bibfnamefont{Y.~P.} \bibnamefont{{Zhang}}},
  \bibinfo{author}{\bibfnamefont{J.~R.} \bibnamefont{{Ensher}}},
  \bibinfo{author}{\bibfnamefont{A.~S.} \bibnamefont{{Estrin}}},
  \bibinfo{author}{\bibfnamefont{C.}~\bibnamefont{{Boisseau}}},
  \bibinfo{author}{\bibfnamefont{R.}~\bibnamefont{{C{\^o}t{\'e}}}},
  \bibinfo{author}{\bibfnamefont{E.~E.} \bibnamefont{{Eyler}}}, and
  \bibinfo{author}{\bibfnamefont{P.~L.} \bibnamefont{{Gould}}},
  \bibinfo{year}{2003}, \bibinfo{journal}{Physical Review Letters}
  \textbf{\bibinfo{volume}{91}}(\bibinfo{number}{18}), \bibinfo{pages}{183002}.

\bibitem[{\citenamefont{Fortagh} \emph{et~al.}(1998)\citenamefont{Fortagh,
  Grossmann, Zimmermann, and H\"{a}nsch}}]{For98}
\bibinfo{author}{\bibnamefont{Fortagh}, \bibfnamefont{J.}},
  \bibinfo{author}{\bibfnamefont{A.}~\bibnamefont{Grossmann}},
  \bibinfo{author}{\bibfnamefont{C.}~\bibnamefont{Zimmermann}}, and
  \bibinfo{author}{\bibfnamefont{T.~W.} \bibnamefont{H\"{a}nsch}},
  \bibinfo{year}{1998}, \bibinfo{journal}{Phys. Rev. Lett.}
  \textbf{\bibinfo{volume}{81}}(\bibinfo{number}{24}), \bibinfo{pages}{5310}.

\bibitem[{\citenamefont{Gallagher}(1994)}]{Gal94}
\bibinfo{author}{\bibnamefont{Gallagher}, \bibfnamefont{T.~F.}},
  \bibinfo{year}{1994}, \emph{\bibinfo{title}{Rydberg Atoms}}
  (\bibinfo{publisher}{Cambridge University Press}).

\bibitem[{\citenamefont{Gupta} \emph{et~al.}(2005)\citenamefont{Gupta, Murch,
  Moore, Purdy, and Stamper-Kurn}}]{gup05}
\bibinfo{author}{\bibnamefont{Gupta}, \bibfnamefont{S.}},
  \bibinfo{author}{\bibfnamefont{K.~W.} \bibnamefont{Murch}},
  \bibinfo{author}{\bibfnamefont{K.~L.} \bibnamefont{Moore}},
  \bibinfo{author}{\bibfnamefont{T.~P.} \bibnamefont{Purdy}}, and
  \bibinfo{author}{\bibfnamefont{D.~M.} \bibnamefont{Stamper-Kurn}},
  \bibinfo{year}{2005}, \bibinfo{journal}{Physical Review Letters}
  \textbf{\bibinfo{volume}{95}}(\bibinfo{number}{14}), \bibinfo{eid}{143201}
  (pages~\bibinfo{numpages}{4}).

\bibitem[{\citenamefont{H\"ansel} \emph{et~al.}(2001)\citenamefont{H\"ansel,
  Hommelhoff, H\"ansch, and Reichel}}]{Hansel01}
\bibinfo{author}{\bibnamefont{H\"ansel}, \bibfnamefont{W.}},
  \bibinfo{author}{\bibfnamefont{P.}~\bibnamefont{Hommelhoff}},
  \bibinfo{author}{\bibfnamefont{T.}~\bibnamefont{H\"ansch}}, and
  \bibinfo{author}{\bibfnamefont{J.}~\bibnamefont{Reichel}},
  \bibinfo{year}{2001}, \bibinfo{journal}{Nature} (\bibinfo{number}{498}),
  \bibinfo{pages}{498}.

\bibitem[{\citenamefont{{Heidemann}}
  \emph{et~al.}(2007)\citenamefont{{Heidemann}, {Raitzsch}, {Bendkowsky},
  {Butscher}, {L{\"o}w}, {Santos}, and {Pfau}}}]{Heidemann07}
\bibinfo{author}{\bibnamefont{{Heidemann}}, \bibfnamefont{R.}},
  \bibinfo{author}{\bibfnamefont{U.}~\bibnamefont{{Raitzsch}}},
  \bibinfo{author}{\bibfnamefont{V.}~\bibnamefont{{Bendkowsky}}},
  \bibinfo{author}{\bibfnamefont{B.}~\bibnamefont{{Butscher}}},
  \bibinfo{author}{\bibfnamefont{R.}~\bibnamefont{{L{\"o}w}}},
  \bibinfo{author}{\bibfnamefont{L.}~\bibnamefont{{Santos}}}, and
  \bibinfo{author}{\bibfnamefont{T.}~\bibnamefont{{Pfau}}},
  \bibinfo{year}{2007}, \bibinfo{journal}{ArXiv Quantum Physics e-prints}
  \eprint{quant-ph/0701120}.

\bibitem[{\citenamefont{Jaksch} \emph{et~al.}(2000)\citenamefont{Jaksch, Cirac,
  Zoller, Rolston, Cote, and Lukin}}]{Jaksch:00}
\bibinfo{author}{\bibnamefont{Jaksch}, \bibfnamefont{D.}},
  \bibinfo{author}{\bibfnamefont{J.~I.} \bibnamefont{Cirac}},
  \bibinfo{author}{\bibfnamefont{P.}~\bibnamefont{Zoller}},
  \bibinfo{author}{\bibfnamefont{S.~L.} \bibnamefont{Rolston}},
  \bibinfo{author}{\bibfnamefont{R.}~\bibnamefont{Cote}}, and
  \bibinfo{author}{\bibfnamefont{M.~D.} \bibnamefont{Lukin}},
  \bibinfo{year}{2000}, \bibinfo{journal}{Phys. Rev. Lett.}
  \textbf{\bibinfo{volume}{85}}, \bibinfo{pages}{2208}.

\bibitem[{\citenamefont{Ketterle}(2002)}]{Ket02}
\bibinfo{author}{\bibnamefont{Ketterle}, \bibfnamefont{W.}},
  \bibinfo{year}{2002}, \bibinfo{journal}{Rev. Mod. Phys.}
  (\bibinfo{number}{74}), \bibinfo{pages}{1131}.

\bibitem[{\citenamefont{Killian} \emph{et~al.}(1999)\citenamefont{Killian,
  Kulin, Bergeson, Orozco, Orzel, and Rolston}}]{Killian:99}
\bibinfo{author}{\bibnamefont{Killian}, \bibfnamefont{T.~C.}},
  \bibinfo{author}{\bibfnamefont{S.}~\bibnamefont{Kulin}},
  \bibinfo{author}{\bibfnamefont{S.~D.} \bibnamefont{Bergeson}},
  \bibinfo{author}{\bibfnamefont{L.~A.} \bibnamefont{Orozco}},
  \bibinfo{author}{\bibfnamefont{C.}~\bibnamefont{Orzel}}, and
  \bibinfo{author}{\bibfnamefont{S.~L.} \bibnamefont{Rolston}},
  \bibinfo{year}{1999}, \bibinfo{journal}{Phys. Rev. Lett.}
  \textbf{\bibinfo{volume}{83}}(\bibinfo{number}{23}), \bibinfo{pages}{4776}.

\bibitem[{\citenamefont{Koch}(1983)}]{Koc83}
\bibinfo{author}{\bibnamefont{Koch}, \bibfnamefont{P.}}, \bibinfo{year}{1983},
  \emph{\bibinfo{title}{Rydberg studies using fast beams}}, Rydberg states of
  atoms and molecules (\bibinfo{publisher}{Cambridge University Press}).

\bibitem[{\citenamefont{Kolokolov and Skrotskii}(1974)}]{Kolokolov:74}
\bibinfo{author}{\bibnamefont{Kolokolov}, \bibfnamefont{A.}}, and
  \bibinfo{author}{\bibfnamefont{G.}~\bibnamefont{Skrotskii}},
  \bibinfo{year}{1974}, \bibinfo{journal}{Opt. Spectrosc.}
  \textbf{\bibinfo{volume}{36}}, \bibinfo{pages}{127}.

\bibitem[{\citenamefont{Lett} \emph{et~al.}(1989)\citenamefont{Lett, Phillips,
  Rolston, Tanner, Watts, and Westbrook}}]{lett:1989}
\bibinfo{author}{\bibnamefont{Lett}, \bibfnamefont{P.}},
  \bibinfo{author}{\bibfnamefont{W.~D.} \bibnamefont{Phillips}},
  \bibinfo{author}{\bibfnamefont{S.~L.} \bibnamefont{Rolston}},
  \bibinfo{author}{\bibfnamefont{C.~E.} \bibnamefont{Tanner}},
  \bibinfo{author}{\bibfnamefont{R.~N.} \bibnamefont{Watts}}, and
  \bibinfo{author}{\bibfnamefont{C.~I.} \bibnamefont{Westbrook}},
  \bibinfo{year}{1989}, \bibinfo{journal}{J. Opt. Soc. Am. B}
  \textbf{\bibinfo{volume}{6}}(\bibinfo{number}{11}), \bibinfo{pages}{2084}.

\bibitem[{\citenamefont{{MacAdam} and {Hwang}}(2003)}]{Adam:03}
\bibinfo{author}{\bibnamefont{{MacAdam}}, \bibfnamefont{K.~B.}}, and
  \bibinfo{author}{\bibfnamefont{C.~S.} \bibnamefont{{Hwang}}},
  \bibinfo{year}{2003}, \bibinfo{journal}{Review of Scientific Instruments}
  \textbf{\bibinfo{volume}{74}}, \bibinfo{pages}{2267}.

\bibitem[{\citenamefont{{Mourachko}}
  \emph{et~al.}(1998)\citenamefont{{Mourachko}, {Comparat}, {de Tomasi},
  {Fioretti}, {Nosbaum}, {Akulin}, and {Pillet}}}]{Mourachko:1998}
\bibinfo{author}{\bibnamefont{{Mourachko}}, \bibfnamefont{I.}},
  \bibinfo{author}{\bibfnamefont{D.}~\bibnamefont{{Comparat}}},
  \bibinfo{author}{\bibfnamefont{F.}~\bibnamefont{{de Tomasi}}},
  \bibinfo{author}{\bibfnamefont{A.}~\bibnamefont{{Fioretti}}},
  \bibinfo{author}{\bibfnamefont{P.}~\bibnamefont{{Nosbaum}}},
  \bibinfo{author}{\bibfnamefont{V.~M.} \bibnamefont{{Akulin}}}, and
  \bibinfo{author}{\bibfnamefont{P.}~\bibnamefont{{Pillet}}},
  \bibinfo{year}{1998}, \bibinfo{journal}{Physical Review Letters}
  \textbf{\bibinfo{volume}{80}}, \bibinfo{pages}{253}.

\bibitem[{\citenamefont{Ottl} \emph{et~al.}(2006)\citenamefont{Ottl, Ritter,
  Kohl, and Esslinger}}]{ottl:063118}
\bibinfo{author}{\bibnamefont{Ottl}, \bibfnamefont{A.}},
  \bibinfo{author}{\bibfnamefont{S.}~\bibnamefont{Ritter}},
  \bibinfo{author}{\bibfnamefont{M.}~\bibnamefont{Kohl}}, and
  \bibinfo{author}{\bibfnamefont{T.}~\bibnamefont{Esslinger}},
  \bibinfo{year}{2006}, \bibinfo{journal}{Review of Scientific Instruments}
  \textbf{\bibinfo{volume}{77}}(\bibinfo{number}{6}), \bibinfo{eid}{063118}
  (pages~\bibinfo{numpages}{18}).

\bibitem[{\citenamefont{Phillips}(1998)}]{Phillips:98}
\bibinfo{author}{\bibnamefont{Phillips}, \bibfnamefont{W.~D.}},
  \bibinfo{year}{1998}, \bibinfo{journal}{Rev. Mod. Phys.}
  \textbf{\bibinfo{volume}{70}}, \bibinfo{pages}{721}.

\bibitem[{\citenamefont{Pritchard}(1983)}]{Pritchard:83}
\bibinfo{author}{\bibnamefont{Pritchard}, \bibfnamefont{D.~E.}},
  \bibinfo{year}{1983}, \bibinfo{journal}{Phys. Rev. Lett.}
  \textbf{\bibinfo{volume}{51}}(\bibinfo{number}{15}), \bibinfo{pages}{1336}.

\bibitem[{\citenamefont{Robert} \emph{et~al.}(2001)\citenamefont{Robert,
  Sirjean, Browaeys, Poupard, Nowak, Boiron, Westbrook, and Aspect}}]{Robert01}
\bibinfo{author}{\bibnamefont{Robert}, \bibfnamefont{A.}},
  \bibinfo{author}{\bibfnamefont{O.}~\bibnamefont{Sirjean}},
  \bibinfo{author}{\bibfnamefont{A.}~\bibnamefont{Browaeys}},
  \bibinfo{author}{\bibfnamefont{J.}~\bibnamefont{Poupard}},
  \bibinfo{author}{\bibfnamefont{S.}~\bibnamefont{Nowak}},
  \bibinfo{author}{\bibfnamefont{D.}~\bibnamefont{Boiron}},
  \bibinfo{author}{\bibfnamefont{C.~I.} \bibnamefont{Westbrook}}, and
  \bibinfo{author}{\bibfnamefont{A.}~\bibnamefont{Aspect}},
  \bibinfo{year}{2001}, \bibinfo{journal}{Science} (\bibinfo{number}{292}),
  \bibinfo{pages}{5516}.

\bibitem[{\citenamefont{{Robinson}}
  \emph{et~al.}(2000)\citenamefont{{Robinson}, {Tolra}, {Noel}, {Gallagher},
  and {Pillet}}}]{Robinson:2000}
\bibinfo{author}{\bibnamefont{{Robinson}}, \bibfnamefont{M.~P.}},
  \bibinfo{author}{\bibfnamefont{B.~L.} \bibnamefont{{Tolra}}},
  \bibinfo{author}{\bibfnamefont{M.~W.} \bibnamefont{{Noel}}},
  \bibinfo{author}{\bibfnamefont{T.~F.} \bibnamefont{{Gallagher}}}, and
  \bibinfo{author}{\bibfnamefont{P.}~\bibnamefont{{Pillet}}},
  \bibinfo{year}{2000}, \bibinfo{journal}{Physical Review Letters}
  \textbf{\bibinfo{volume}{85}}, \bibinfo{pages}{4466}.

\bibitem[{\citenamefont{{Singer}}
  \emph{et~al.}(2004{\natexlab{a}})\citenamefont{{Singer}, {Reetz-Lamour},
  {Amthor}, {Marcassa}, and {Weidem{\"u}ller}}}]{Singer:04}
\bibinfo{author}{\bibnamefont{{Singer}}, \bibfnamefont{K.}},
  \bibinfo{author}{\bibfnamefont{M.}~\bibnamefont{{Reetz-Lamour}}},
  \bibinfo{author}{\bibfnamefont{T.}~\bibnamefont{{Amthor}}},
  \bibinfo{author}{\bibfnamefont{L.~G.} \bibnamefont{{Marcassa}}}, and
  \bibinfo{author}{\bibfnamefont{M.}~\bibnamefont{{Weidem{\"u}ller}}},
  \bibinfo{year}{2004}{\natexlab{a}}, \bibinfo{journal}{Physical Review
  Letters} \textbf{\bibinfo{volume}{93}}(\bibinfo{number}{16}),
  \bibinfo{pages}{163001}.

\bibitem[{\citenamefont{{Singer}}
  \emph{et~al.}(2004{\natexlab{b}})\citenamefont{{Singer}, {Reetz-Lamour},
  {Amthor}, {Marcassa}, and {Weidem{\"u}ller}}}]{Singer:2004}
\bibinfo{author}{\bibnamefont{{Singer}}, \bibfnamefont{K.}},
  \bibinfo{author}{\bibfnamefont{M.}~\bibnamefont{{Reetz-Lamour}}},
  \bibinfo{author}{\bibfnamefont{T.}~\bibnamefont{{Amthor}}},
  \bibinfo{author}{\bibfnamefont{L.~G.} \bibnamefont{{Marcassa}}}, and
  \bibinfo{author}{\bibfnamefont{M.}~\bibnamefont{{Weidem{\"u}ller}}},
  \bibinfo{year}{2004}{\natexlab{b}}, \bibinfo{journal}{Physical Review
  Letters} \textbf{\bibinfo{volume}{93}}(\bibinfo{number}{16}),
  \bibinfo{pages}{163001}.

\bibitem[{\citenamefont{Streed} \emph{et~al.}(2006)\citenamefont{Streed,
  Chikkatur, Gustavson, Boyd, Torii, Schneble, Campbell, Pritchard, and
  Ketterle}}]{streed:023106}
\bibinfo{author}{\bibnamefont{Streed}, \bibfnamefont{E.~W.}},
  \bibinfo{author}{\bibfnamefont{A.~P.} \bibnamefont{Chikkatur}},
  \bibinfo{author}{\bibfnamefont{T.~L.} \bibnamefont{Gustavson}},
  \bibinfo{author}{\bibfnamefont{M.}~\bibnamefont{Boyd}},
  \bibinfo{author}{\bibfnamefont{Y.}~\bibnamefont{Torii}},
  \bibinfo{author}{\bibfnamefont{D.}~\bibnamefont{Schneble}},
  \bibinfo{author}{\bibfnamefont{G.~K.} \bibnamefont{Campbell}},
  \bibinfo{author}{\bibfnamefont{D.~E.} \bibnamefont{Pritchard}}, and
  \bibinfo{author}{\bibfnamefont{W.}~\bibnamefont{Ketterle}},
  \bibinfo{year}{2006}, \bibinfo{journal}{Review of Scientific Instruments}
  \textbf{\bibinfo{volume}{77}}(\bibinfo{number}{2}), \bibinfo{eid}{023106}
  (pages~\bibinfo{numpages}{13}).

\bibitem[{\citenamefont{{Tong}} \emph{et~al.}(2004)\citenamefont{{Tong},
  {Farooqi}, {Stanojevic}, {Krishnan}, {Zhang}, {C{\^o}t{\'e}}, {Eyler}, and
  {Gould}}}]{Tong:2004}
\bibinfo{author}{\bibnamefont{{Tong}}, \bibfnamefont{D.}},
  \bibinfo{author}{\bibfnamefont{S.~M.} \bibnamefont{{Farooqi}}},
  \bibinfo{author}{\bibfnamefont{J.}~\bibnamefont{{Stanojevic}}},
  \bibinfo{author}{\bibfnamefont{S.}~\bibnamefont{{Krishnan}}},
  \bibinfo{author}{\bibfnamefont{Y.~P.} \bibnamefont{{Zhang}}},
  \bibinfo{author}{\bibfnamefont{R.}~\bibnamefont{{C{\^o}t{\'e}}}},
  \bibinfo{author}{\bibfnamefont{E.~E.} \bibnamefont{{Eyler}}}, and
  \bibinfo{author}{\bibfnamefont{P.~L.} \bibnamefont{{Gould}}},
  \bibinfo{year}{2004}, \bibinfo{journal}{Physical Review Letters}
  \textbf{\bibinfo{volume}{93}}(\bibinfo{number}{6}), \bibinfo{pages}{063001}.

\bibitem[{\citenamefont{{Vogt}} \emph{et~al.}(2006)\citenamefont{{Vogt},
  {Viteau}, {Zhao}, {Chotia}, {Comparat}, and {Pillet}}}]{Vogt:2006}
\bibinfo{author}{\bibnamefont{{Vogt}}, \bibfnamefont{T.}},
  \bibinfo{author}{\bibfnamefont{M.}~\bibnamefont{{Viteau}}},
  \bibinfo{author}{\bibfnamefont{J.}~\bibnamefont{{Zhao}}},
  \bibinfo{author}{\bibfnamefont{A.}~\bibnamefont{{Chotia}}},
  \bibinfo{author}{\bibfnamefont{D.}~\bibnamefont{{Comparat}}}, and
  \bibinfo{author}{\bibfnamefont{P.}~\bibnamefont{{Pillet}}},
  \bibinfo{year}{2006}, \bibinfo{journal}{Physical Review Letters}
  \textbf{\bibinfo{volume}{97}}(\bibinfo{number}{8}), \bibinfo{pages}{083003}.

\end{thebibliography}
\end{document}